\documentclass[twocolumn,prl]{revtex4}
\usepackage{float}
\usepackage{makeidx}
\usepackage{graphicx,amsmath}
\usepackage{CJK}

\begin{document}
\begin{CJK*}{GB}{gbsn}

\title{Momentum-resolved Raman spectroscopy of bound molecules in ultracold Fermi gas}
\author{Zhengkun Fu, Pengjun Wang, Lianghui Huang, Zengming Meng, Jing Zhang$^{\dagger}$}
\affiliation{State Key Laboratory of Quantum Optics and Quantum
Optics Devices, Institute of Opto-Electronics, Shanxi University,
Taiyuan 030006, P.R.China \label{in}}

\begin{abstract}

The binding energy of Feshbach molecules from a two component Fermi
gas of $^{40}$K atoms has been experimentally measured with the
momentum-resolved Raman spectroscopy. Comparing with the
radio-frequency spectroscopy, in the present experiment the signal
of unpaired (free atoms) and the bound molecules can be directly
observed and the binding energy can be simultaneously determined in
a single running experiment. The energy-momentum dispersion spectra
of the strongly interacting ultracold Fermi gas in BEC side are also
measured and reconstructed. The present experimental technology of
the momentum-resolved Raman spectroscopy can be easily extended to
perform spatially momentum-resolved Raman spectroscopy and to obtain
the response spectra of a homogeneous system in the local density
approximation.

\end{abstract}

\maketitle
\end{CJK*}

Interacting Fermi gas \cite{Giorgini}, is a simple, clean and easy
controllable system with rich physical behaviors, which provides a
new platform for studying the frustrating problems in the condensed
matter physics and the quantum simulation of many-body system
\cite{Bloch}, such as high-temperature superfluidity and BEC-BCS
crossover. Magnetic field induced Feshbach resonances present a
means to precisely control the interaction of ultracold Fermi atoms
in different spin states over several orders of magnitude
\cite{one0,one1,one2,one3}. This method allows us to create
molecules \cite{one4,one5} and fermionic superfluidity in balanced
and unbalanced components \cite{one6,one7,one8}. Radio-frequency
(RF) spectroscopy has become a powerful tool for studying
single-particle excitations in degenerate Fermi gases, such as to
probe the pair size \cite{one9}, energy excitation spectrum
\cite{one10,one11,one11-1,one11-2}, and paring gap
\cite{one12,one13,one14}. An other important tool in this field is
Bragg spectroscopy, which has been used for probing density-density
correlations \cite{one15} and measuring the universal contact of an
interacting Fermi gas \cite{one16}. There are also many other
detection methods of Fermi pairing proposed theoretically such as
Stokes-scattering method \cite{method1}, interferometric method
\cite{method2}, electromagnetically induced transparency method
\cite{method3}.

The Raman spectroscopy technology \cite{one17} has been proposed to
probe the one-particle excitation in degenerate Fermi gases. In the
Raman process, the atoms in initial state absorb a photon from a
Raman laser beam and immediately emit a photon into another Raman
laser beam. During this process the atoms are transferred into an
other internal state with different momentum. Compared to the RF
spectrum, the Raman spectroscopy offers several advantages, e.g.,
spatial selectivity, tunability of transferred momentum and weak
sensitivity to final state interaction \cite{one18}. Recently, we
have experimentally measured the dispersion of a non-interacting
degenerate Fermi gas by means of the momentum-resolved Raman
spectroscopy technology \cite{one19}.

In this letter, we present the first experimental achievement on
exploiting Raman spectroscopy to probe Feshbach molecules in
ultracold Fermi gas. The Feshbach molecules are created in $^{40}$K
atomic gas consisting of an equal mixture of atoms in the
$|F=9/2,m_{F}=-9/2\rangle$ and $|F=9/2,m_{F}=-7/2\rangle$ states by
ramping the magnetic field from above the s-wave Feshbach resonance
202.2 G to a value below the resonance. We apply a Gaussian shape
pulse of Raman laser to transfer atoms from the state
$|F=9/2,m_{F}=-7/2\rangle$ to the final state
$|F=9/2,m_{F}=-5/2\rangle$ and record the momentum distribution of
the atoms in $|F=9/2,m_{F}=-5/2\rangle$ state by time-of-flight
(TOF) absorption image. We observe simultaneously the composition of
gas, including unpaired atoms and bound molecules in BEC side and
determine the binding energy by the momentum distribution of the
atoms in $|F=9/2,m_{F}=-5/2\rangle$ state in a single running
experiment. The momentum distributions of the atoms in
$|F=9/2,m_{F}=-5/2\rangle$ state are recorded as a function of the
frequency difference between the two Raman lasers, and then the
spectral function of the ultracold Fermi gas is reconstructed.

\begin{figure}
\centerline{
\includegraphics[width=3.2in]{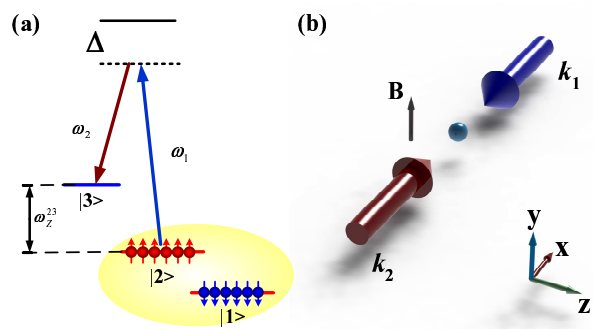}
} \vspace{0.1in}
\caption{(Color online). \textbf{Geometry and energy level diagram
for Raman spectroscopy}. \textbf{a,} The atomic level diagram for
Raman transition from the state $|2\rangle$ to $|3\rangle$ and the
bound molecules formed by the atoms in states $|1\rangle$ and
$|2\rangle$. \textbf{b,} Schematic of the Raman spectroscopy
experiment. Two Raman beams counterpropagating along $\pm\hat{x}$
illuminate the ultracold Fermi gas and the black arrow indicates the
direction of the bias magnetic field $B$. \label{Fig1} }
\end{figure}

We consider the atomic system with three internal states labeled by
$|1\rangle, |2\rangle$ and $|3\rangle$, as shown in Fig. 1 (a).
There is a interaction between the atoms in state $|1\rangle$ and
$|2\rangle$, which forms the molecules with the binding energy
$E_{b}$, and the state $|3\rangle$ is non-interacting with both
states $|2\rangle$ and $|1\rangle$. The Raman laser
frequency-difference is chosen to transfer the atoms in state
$|2\rangle$ to $|3\rangle$, but which is not in resonance with any
transition. For the Raman process, we define the effective Raman
coupling as $\Omega=\Omega_{1}\Omega_{2}/\Delta$, here $\Omega_{i}$
is the Rabi frequency of laser beam $i$ with frequency $\omega_{i}$
and wave vector $\textbf{k}_{i}$, $\Delta$ is the detuning between
Raman laser field and intermediated excited states. According to
energy and momentum conservation, one obtains the momentum transfer
$\textbf{q}_{r}=\textbf{k}_{1}-\textbf{k}_{2}$ and the energy shift
$\hbar\triangle\omega=\hbar(\omega_{1}-\omega_{2})$ of the atoms and
it can also be described by expressions of the outgoing channel
\begin{eqnarray}
\hbar\triangle\omega=E_{Z}^{32}(B)-\epsilon_{|2\rangle}^{Initial}(\textbf{k})+\epsilon_{|3\rangle}^{Final}(\textbf{k}+\textbf{q}_{r}).
\end{eqnarray}
Here $E_{Z}^{32}(B)$ is the Zeeman energy split between the state
$|3\rangle$ and $|2\rangle$ in a magnetic field $\textbf{B}$.
$\epsilon_{|3\rangle}^{Final}(\textbf{k}+\textbf{q}_{r})$ is the
energy momentum dispersion of atoms in final state $|3\rangle$. The
spin flipped atoms have only very weak interactions with the other
atoms which means that the dispersion can be expressed with the
usual free-particle dispersion:
$\epsilon_{|3\rangle}^{Final}(\textbf{k}+\textbf{q}_{r})=\hbar^{2}(|\textbf{k}+\textbf{q}_{r}|)^{2}/2m$,
where $m$ is the atomic mass.
$\epsilon_{|2\rangle}^{Initial}(\textbf{k})$ stands for the
dispersion of atoms in state $|2\rangle$ strongly interacting with
the state $|1\rangle$, which can be reconstructed by the momentum
distributions of the atoms in $|3\rangle$ state as a function of the
frequency difference of the two Raman lasers,
\begin{eqnarray}
\epsilon_{|2\rangle}^{Initial}(\textbf{k})=E_{Z}^{32}(B)-\hbar\triangle\omega+\hbar^{2}(|\textbf{k}+\textbf{q}_{r}|)^{2}/2m.
\end{eqnarray}

The apparatus and the basic experimental methods have been described
in our previous publications \cite{four2, four3, four4, four5,
four6}, in which the Bose-Fermi mixtures are cooled in magnetic
field and transported into an optical trap. The experiment begin
with a degenerate Fermi gas about $2\times10^{6}$ $^{40}$K in the
$|F=9/2,m_{F}=9/2\rangle$ internal state, which has been
evaporatively cooled to $T/T_{F}\approx0.3$ with bosonic $^{87}$Rb
atoms inside the crossed optical trap, where $T$ is the temperature,
$T_{F}$ is the Fermi temperature defined by
$T_{F}=E_{F}/k_{B}=\frac{(\hbar\overline{\omega})}{k_{B}}(6
N)^{1/3}$, and $\overline{\omega}$ is the geometric mean trapping
frequency, $N$ is the number of fermions. A 780 $nm$ laser pulse of
0.03 $ms$ is used to remove the $^{87}$Rb atoms in the mixture
without heating of $^{40}$K atoms. Subsequently, the fermionic atoms
are transferred into the lowest state $|F=9/2,m_{F}=-9/2\rangle$ via
a rapid adiabatic passage induced by a radio frequency field of 80
$ms$ at 4 $G$. In order to prepare the Fermi gas $^{40}$K in an
equal mixture of $|F=9/2,m_{F}=-9/2\rangle$ (regarded as
$|1\rangle$) and $|F=9/2,m_{F}=-7/2\rangle$ (regarded as
$|2\rangle$) states, a homogeneous bias magnetic field produced by
the quadrupole coils (operating in the Helmholtz configuration) is
raised to about $B\approx219.4$ $G$ in the $\hat{y}$ direction and
then a radio frequency ramp around 47.45 $MHz$ is applied for 50
$ms$.

It is crucial in the experiment to control the high magnetic field
precisely and reduce the drift and AC field noise. The current
through the coils is controlled by the external regulator relying on
a precision current transducer (Danfysik ultastable 867-60I). An
analog current signal is generated from the current transducer which
is proportional to the primary current. The current signal is
transformed into the voltage signal by a precision resistor with a
low temperature coefficient of $15$ $ppm/^{0}C$. Then a regulator
compares this current value with a given voltage value from a
computer. The output error signal from the regulator actively
stabilizes the current with the PID
(proportional-integral-derivative) controller acting on the MOSFET
(metal-oxide-semiconductor field-effect transistor). In order to
reduce the current noise and decouple the control circuit from the
main current, a conventional battery is used to power the circuit.

In order to create molecules, the homogeneous bias magnetic field is
ramped from 204 $G$ to a value B below the Feshbach resonance
located at 202.2 $G$ at a rate of about 0.08 $G/ms$. This procedure
would therefore result in a mixture of weakly bound molecules and
unpaired atoms in $|F=9/2,m_{F}=-9/2\rangle$ and
$|F=9/2,m_{F}=-7/2\rangle$ states. To probe the binding energy of
molecules by Raman spectroscopy technology, a pair of $773$ $nm$
Raman laser from a Ti-sapphire laser with the frequency difference
$\Delta\omega$, counterpropagating along the $\hat{x}$ axis, which
couple the two hyperfine states $|F=9/2,m_{F}=-7/2\rangle$
($|2\rangle$) and $|F=9/2,m_{F}=-5/2\rangle$ ($|3\rangle$) as shown
in Fig. 1(b). The momentum transferred to atoms during the Raman
process is $|\textbf{q}_{r}|=2k_{r}\sin(\theta/2)$, where
$k_{r}=2\pi /\lambda$ is the single-photon recoil momentum,
$\lambda$ is the wavelength of the Raman beam, and $\theta=180^{o}$
is the intersecting angle of two Raman beams. Here, $\hbar k_{r}$
and $E_{r}=(\hbar k_{r})^{2}/2m = h\times 8.34$ $kHz$ are the
defined units of momentum and energy for the later figures. Two
Raman beams are frequency-shifted -75 $MHz$ and -122 $MHz$ by two
single-pass acousto-optic modulators (AOM) respectively. In this way
the relative frequency difference between the two laser beams is
precisely controlled by the two signal generators used for two AOMs.
After out of the optical fibers, the two Raman beams with intensity
$I= 50$ $mW$, are intersecting in the atomic cloud with $1/e^{2}$
radii of 200 $\mu m$ and are linearly polarized along $\hat{z}$ and
$\hat{y}$ axis directions respectively, which correspond to $\pi$
and $\sigma$ of the quantization axis $\hat{y}$. A Gaussian shape
pulse of the Raman laser with a duration time about 70 $\mu s$ is
applied to transfer atoms from the initial state $|9/2,-7/2\rangle$
to the final state $|9/2,-5/2\rangle$. The pulse is generated by the
voltage-controlled RF attenuators of AOMs' driver. The Gaussian
envelop hence results in the elimination of the side lobes in Raman
spectra. Here, when the atoms are in Raman resonance (at about 47.2
$MHz$) between $|F=9/2,m_{F}=-7/2\rangle$ and
$|F=9/2,m_{F}=-5/2\rangle$ near Feshbach resonance (202.2 $G$), the
energy split between $|F=9/2,m_{F}=-9/2\rangle$ and
$|F=9/2,m_{F}=-7/2\rangle$ is about 44.8 $MHz$, which is a very
large detuning for Raman transition. After the Raman laser pulse, we
abruptly turn off the optical trap and the magnetic field, and let
the atoms ballistically expand for 12 $ms$ in a magnetic field
gradient applied along $\hat{z}$ and take TOF absorption image along
$\hat{y}$. The atoms in different hyperfine states and momentum
states are spatially separated and thus can be easily analyzed.

\begin{figure}
\centerline{
\includegraphics[width=3in]{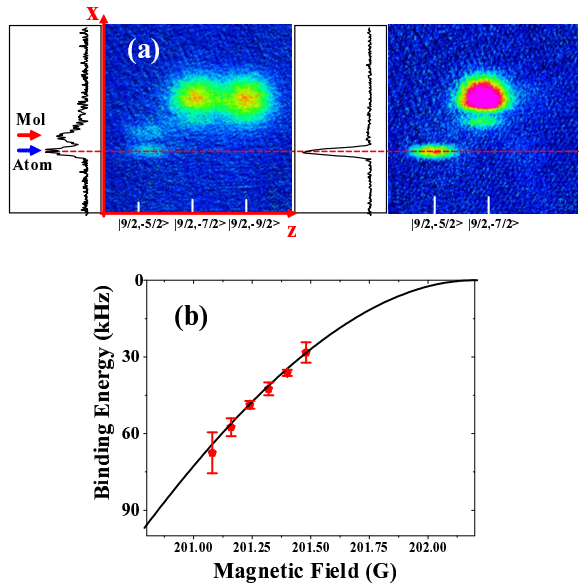}
} \vspace{0.1in}
\caption{(Color online). \textbf{Determining the binding energy of
s-wave molecules. } \textbf{a,} Absorption images of Fermi gas and
integrated optical density of atoms in $|F=9/2,m_{F}=-5/2\rangle$
state along $\hat{z}$ with the mixture and the single spin state
$|F=9/2,m_{F}=-7/2\rangle$. The red dashed line indicates the same
position in $\hat{x}$ for free atoms. Mol: molecules. \textbf{b,}
The measured binding energies of molecules on the BEC side of the
s-wave Feshbach resonance are plotted versus magnetic fields (red
pentagon). The solid line is the theoretical calculation of the
binding energy of molecules for $^{40}K$ atoms in
$|F=9/2,m_{F}=-9/2\rangle$ and $|F=9/2,m_{F}=-7/2\rangle$ states.
\label{Fig2} }
\end{figure}

We first measure the function of the binding energy of molecules
$E_{b}(B)$ versus the magnetic field on the BEC side of Feshbach
resonance. By selecting the appropriate frequency difference
$\Delta\omega$ of the two Raman beams, the two distinctly different
momentum distributions of the atoms in $|9/2,-5/2\rangle$ state
along $\hat{x}$ appear in a TOF absorption image as shown in Fig.
2(a). The phenomenon is induced by the free atoms and bond molecules
with the different momentums, which satisfies Eq. 1 simultaneously
when fixing frequency difference $\Delta\omega$ of two Raman beams.
In order to recognize the different momentum distributions resulting
from the free atoms or the bond molecules, we perform the same
procedures and the only difference is to prepare atoms in the single
spin state $|F=9/2,m_{F}=-7/2\rangle$ instead of the equal mixture.
The TOF image only shows a single momentum distribution of the atoms
in $|9/2,-5/2\rangle$ state, which corresponds to the free atoms as
shown in Fig. 2(a). The peak corresponding to unpair atoms is
narrower and exhibits symmetric shape. The asymmetric peak
corresponds to the dissociation of the bound molecules. This result
shows that the composition of gas includes the unpaired atoms and
bound molecules in BEC side. To determine the binding energy of
molecules exactly, we first determine the maximum molecule signal
and then fit this curve to extract the distance x of two peaks of
the atoms in $|9/2,-5/2\rangle$ state from the atom optical density
integration along $\hat{z}$ direction. The relationship between the
binding energy of molecules and the distance x of two peaks can be
determined by preparing atoms in the single spin state
$|F=9/2,m_{F}=-7/2\rangle$ and performing respectively two
measurements under two different frequency differences between two
Raman beams, $\Delta\omega_{1}$ and $\Delta\omega_{2}$. The distance
between the atomic momentum distribution peaks measured at the two
cases' of $\Delta\omega_{1}$ and $\Delta\omega_{2}$ corresponds to
the energy $\hbar(\Delta\omega_{1}-\Delta\omega_{2})$. The function
of the molecule binding energy versus the magnetic-field intensity
is plotted in Fig. 2(b). Due to the limit of the momentum
distribution of the unpaired atoms and bound molecules, the binding
energy of molecules can be directly measured in a single running
experiment if the binding energy of molecules is smaller than about
70 $kHz$. When the binding energy of molecules is larger than about
70 $kHz$, we must perform two measurements to determine the
positions of molecules and unpair atoms respectively. We compare the
measured binding energies near the Feshbach resonance with the black
line based on a theoretical calculation of a approximate analytic
formula with the requirement of $a\gg r_{0}$ \cite{one4}
\begin{eqnarray}
E_{b}=\frac{\hbar^{2}}{m(a-r_{0})^{2}},
\end{eqnarray}
where $a$ is the scattering length describing the interaction of
atoms in states $|F=9/2,m_{F}=-7/2\rangle$ and
$|F=9/2,m_{F}=-9/2\rangle$, and $r_{0}\approx60a_{0}$ is the range
of Van der Waals potential \cite{one4}. The experimental results are
consistent with the expected behavior under changing the magnetic
field.

\begin{figure}
\centerline{
\includegraphics[width=3in]{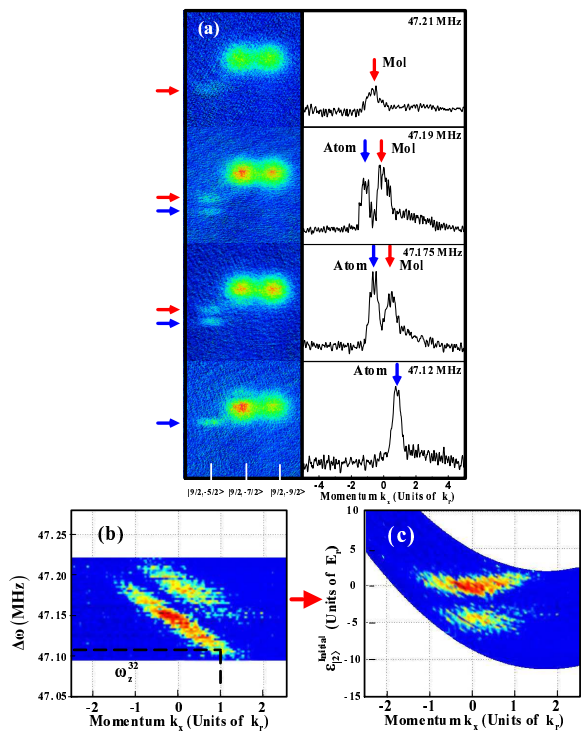}
} \vspace{0.1in}
\caption{(Color online). \textbf{TOF images and the energy-momentum
dispersion reconstructed by momentum-resolved Raman spectroscopy for
ultracold Fermi gas}. \textbf{a,} Absorption images of Fermi mixture
and integrated optical density of atoms in
$|F=9/2,m_{F}=-5/2\rangle$ state at 201.5 $G$ with Raman frequency
difference $\Delta\omega$ from 47.1 $MHz$ to 47.22 $MHz$.
\textbf{b,} The plot is intensity map of the atoms in
$|F=9/2,m_{F}=-5/2\rangle$ state in the ($\Delta\omega, k_{x}$)
plane. The atomic density is displayed with the pseudocolor. The
blue shade regions correspond to the atomic density of zero.
\textbf{c,} The translated intensity spectrum shows the atomic
number of unpaired atoms and bound molecules as a function of the
single particle energy (normalized to $E_{r}$) and momentum $k_{x}$
(normalized to $k_{r}$). \label{Fig3} }
\end{figure}

Now we reconstruct the spectral function of the Fermi gas on the BEC
side by the momentum-resolved Raman spectroscopy. By fixing the
magnetic field and changing the frequency difference between the two
Raman lasers with a step of 2.5 $kHz$, we obtain the TOF absorption
images as function of the frequency difference of Raman lasers as
shown in Fig. 3(a). Then we integrate atomic optical density of the
atoms in $|9/2,-5/2\rangle$ state along $\hat{z}$ direction for each
image and obtain the momentum distributions in $\hat{x}$. All
momentum distributions in $\hat{x}$ of the spin state
$|9/2,-5/2\rangle$ for different frequency differences of the Raman
lasers are plotted into the ($\Delta\omega, k_{x}$) plane, as shown
in Fig. 3(b). According to Eq. 2 and the quadratic energy-momentum
dispersion of the final state, the energy-momentum dispersion of the
initial state (Fig. 3(c)) is obtained from the measured spectrum
(Fig. 3(b)). The reconstructed energy-momentum dispersion
simultaneously presents the characteristics of the unpaired atoms
and bound molecules in BEC side, in which the unpaired atoms show
nearly quadratic dispersion, however the bound molecules does not
exhibit the exactly quadratic dispersion. The bound molecules are
broken into two free atoms with equal and opposite momentum in the
molecule center-of-mass frame by the Raman laser. Thus there is the
extended momentum distribution at the lower negative energy below
the binding energy. The broader dispersion of the bound molecules is
shown in Fig. 3c. In Ref. \cite{one10,one11}, an inverse Abel
transform is used to reconstruct the three-dimensional momentum
distribution when consider the momentum distribution is isotropic.
Then the dispersion spectra along the radial direction of momentum
are obtained. Here, we give the dispersion in $k_{x}$ direction
after integrating one direction of the absorption image. Although
the dispersion spectra between our work and Ref. \cite{one10,one11}
are plotted with the different forms, the dispersion spectra of the
bound molecules show the similar features since they come from the
same wave-function of bound molecules.

In conclusion, we experimentally study the Raman spectroscopy of
ultracold Fermi gas on the BEC side near a s-wave Feshbach
resonance. Since Raman process will transfer a large momentum into
atoms, Raman spectroscopy presents the inherent momentum-resolved
characteristics. The binding energy of molecules can directly be
measured in a single running experiment by means of Raman
spectroscopy. The spectral function of the ultracold Fermi gas is
reconstructed, which presents the different characteristics of the
unpaired atoms and bound molecules in BEC side. This experimental
technology has some advantages compared to RF and Bragg technology
and can be easily extended to perform spatially momentum-resolved
Raman spectroscopy and to obtain the spectra function for the
response of a homogeneous system in the local density approximation.

\begin{acknowledgments}

$^{\dagger}$Corresponding author email: jzhang74@yahoo.com,
jzhang74@sxu.edu.cn

J. Zhang would like to thank Hui Hu for helpful discussions. This
research is supported by National Basic Research Program of China
(Grant No. 2011CB921601), NSFC Project for Excellent Research Team
(Grant No. 61121064), Doctoral Program Foundation of Ministry of
Education China (Grant No. 20111401130001).

\end{acknowledgments}

\end{document}